\documentclass[showpacs,pra]{revtex4}

\usepackage[citecolor=blue,colorlinks=true,linkcolor=blue]{hyperref}
\usepackage{amsmath,amsfonts}
\usepackage{dcolumn}% Align table columns on decimal point

\begin{document}

\title{The leading term of the He$-\bar{p}\mbox{He}^+$ long-range interaction}

\author{Vladimir I. Korobov}
\affiliation{Bogoliubov Laboratory of Theoretical Physics, Joint Institute for Nuclear Research, 141980 Dubna, Russia}

\author{Zhen-Xiang Zhong}
\email{zxzhong@wipm.ac.cn}
\author{Quan-Long Tian}

\affiliation{Division of Theoretical and Interdisciplinary Research,
State Key Laboratory of Magnetic Resonance and Atomic and Molecular Physics,
Wuhan Institute of Physics and Mathematics,
Chinese Academy of Sciences, Wuhan 430071, China}

\date{\today}

\begin{abstract}
The long range interaction between an antiprotonic helium atom $\bar{p}$He$^+$ and helium atom in its ground state is studied. We calculate the dispersion coefficients $C_6$ using the Complex Coordinate Rotation (CCR) formalism in order to comply with the resonant nature of metastable states of the antiprotonic helium. We present as well numerical data on static dipole polarizabilities of antiprotonic helium states. The obtained coefficients $C_6$ may be used to estimate the collisional shift and broadening of transition lines in a low density precision spectroscopy of the antiprotonic helium.
\end{abstract}


\maketitle

\section{Introduction}

In 1991 in experiment at KEK it was discovered that some fraction of antiprotons in a helium target survive for unexpectedly long time \cite{Iwasaki1991}. Later in a series of experiments at CERN  \cite{Yamazaki2002} it has been shown that such antiprotons form an exotic atom, ``antiprotonic helium'', or $\bar{p}$He$^+$. An antiproton stopped in the target is then captured into some atomic state in a helium atom via a charge exchange reaction replacing one of the two electrons. The major part of such states disappears promptly due to annihilation of the antiproton on a nucleus, while a small fraction still survives making up a set of metastable states, which decay predominantly via slow radiative transitions.

Further precision studies of these atoms by laser spectroscopy \cite{Morita1994,Maas1995,Torii1999} reveal reach possibility to investigate various properties of the antiprotonic helium atoms as well as to infer precise data on antiproton \cite{Hayano2007,Hori2013}. Particularly, it provides a stringent test of CPT invariance in a barion sector. Otherwise, assuming validity of CPT invariance one may extract (anti)proton-to-electron mass ratio \cite{Hori2006,Hori2011}. The latter data have been used in the CODATA adjustment of fundamental physical constants \cite{CODATA10}.

Antiprotonic helium can be described as a three particle atomic Coulomb system composed of a helium nucleus, an electron in the $1s$-ground state, and an antiproton occupying a nearly circular orbital with the principal quantum number $n\sim n_0=\sqrt{M^*/m_e}\sim38$, where $M^*$ is the reduced mass of the $\bar{p}$-He pair. Under these conditions the Auger decay is suppressed and this small fraction of states may survive as long as few microseconds even in a liquid helium target.

This longevity has allowed to perform precision spectroscopy of multiple transitions in these atoms with precision which has been rapidly improved from several ppm (parts per million) in the 1990s to a ppb level \cite{Hori2001,Hori2003,Hori2006,Hori2011}, thus became sensitive to the antiproton-to-electron mass ratio ($m_{\bar{p}}/m_e$). Compared with theoretical frequencies of three-body QED calculations \cite{Elander1997,Korobov1997,Kino1999,Korobov2003,Korobov2008} for some selected transitions in $\bar{p}^3$He$^+$ and $\bar{p}^4$He$^+$, these measurement can be used to determine $m_{\bar{p}}/m_e$ \cite{Hayano2007}. The latest determination of the mass ratio is $m_{\bar{p}}/m_e\!=\!1\,836.152\,673\,6(23)$ \cite{Hori2011} which was carried out using the two-photon Doppler reduced laser spectroscopy. At this level of precision the collisional effects as well as Stark effects become important contributions to the total experimental error (see Table 2 in Ref. \cite{Hori2011}).

The density shift and broadening were measured for antiprotonic helium spectral lines with $T=6$ K \cite{Torii1999,Hori2001,Hayano2007}. Good qualitative description of the experimental data were obtained in \cite{Korenman1999} using effective model potential of $\bar{p}$He$^+$$-$He long-range attraction and short-range repulsion interactions. Quantitative agreement with the experimental data was achieved in \cite{Bakalov2000} by direct \emph{ab initio} calculation of an interatomic potential in the Born-Oppenheimer approximation.

Recently ASACUSA collaboration at CERN \cite{Hayano2013} announced that they want to slow down the pulsed antiproton beam up to 150 eV and to cool down the experimental target cell to $T\le1.5$ K. For precision spectroscopy beyond 1 ppb level the long range interaction between antiprotonic helium and ground state helium atoms is of great importance for proper evaluation of collisional effects and their influence on the experimentally observed spectral lines.

In our work we intend to calculate static dipole polarizabilities for metastable states in $\bar{p}^3$He$^+$ and $\bar{p}^4$He$^+$ atoms. We use the Complex Coordinate Rotation (CCR) formalism \cite{Ho1983}, which allows to take into account in a proper way the resonant nature of the metastable states in the antiprotonic helium. Finally, the dispersion coefficients $C_6$ will be evaluated numerically using the same CCR formalism. These coefficients determine the leading contribution to the long-range He-$\bar{p}$He$^+$ interaction. Here we assume that all helium atoms of the target are in the ground state.

Atomic units $(\hbar=e=m_e=1)$ are used throughout this paper.

\section{Theory}

\subsection{Wave functions of He and $\bar{p}$He$^+$}

Both atoms, usual helium and antiprotonic helium, are three-body systems with Coulomb interaction and will be considered using the same variational expansion. The strong interaction between $\bar{p}$ and helium nucleus is strongly suppressed by the centrifugal barrier (the angular momentum of an antiprotonic orbital $l\approx34$) and may be completely neglected.

The nonrelativistic Hamiltonian of a three-body system is taken in a form
\begin{equation}
H =
   -\frac{1}{2\mu_1}\nabla^2_{r_1}
   -\frac{1}{2\mu_2}\nabla^2_{r_2}
   -\frac{1}{M}\nabla_{r_1}\cdot\nabla_{r_2}
   -\frac{Z}{r_1}-\frac{Z}{r_2}+\frac{1}{r_{12}}\,,
\end{equation}
where $\mathbf{r}_1$ and $\mathbf{r}_2$ are position vectors for two negative particles, $\mathbf{r}_{12}=\mathbf{r}_1-\mathbf{r}_2$, $\mu_1=Mm_1/(M+m_1)$ and $\mu_2=Mm_2/(M+m_2)$ are reduced masses, and $M$ is a mass of helium nucleus, the nucleus charge is $Z=2$. We assume that in case of helium atom $m_1=m_2=1$ are masses of electrons, while for the antiprotonic helium we set $m_1=m_{\bar{p}}$ and $m_2=1$, where $m_{\bar{p}}\,$ is a mass of an antiproton.

The helium atoms are in its ground state. Antiprotonic helium is a more complicate object. It presents a quasi adiabatic system with a heavy antiproton orbiting over helium nucleus with a velocity of about 40 times slower then a remaining electron. Using atomic terminology, the electron occupies its ground state: $\psi_{1s}$, while the antiproton may be approximately described by its principal and orbital quantum numbers, $n$ and $l$. Due to interaction between electron and antiproton these quantum numbers are not exact and the wave function is determined by the total angular orbital momentum $L$ and the excitation (or vibrational) quantum number $v$, which are related to the atomic one as follows: $L = l$, $v = n-l-1$.

In our calculations we use a variational expansion based on exponentials with randomly generated parameters. The wave functions both for initial states and for intermediate states (for the second order perturbation calculations) are taken in the form
\begin{equation}
\Psi_L(l_1,l_2)=\sum_{k=1}^{\infty}
\Bigl\{
   U_k{\rm{Re}}[e^{-\alpha_kr_1-\beta_kr_2-\gamma_kr_{12}}]
   +W_k{\rm{Im}}[e^{-\alpha_kr_1-\beta_kr_2-\gamma_kr_{12}}]\Bigr\}
   \mathcal{Y}^{l_1,l_2}_{LM}(\hat{\bf{r}}_1,\hat{\bf{r}}_2)\,,
\end{equation}
where ${\cal{Y}}^{l_1,l_2}_{LM}(\hat{\bf{r}}_1,\hat{\bf{r}}_2)$ are the solid bipolar harmonics as defined in Ref. \cite{Varsha1988}, and $L$ is the total orbital angular momentum of a state. Complex parameters $\alpha_k$, $\beta_k$, and $\gamma_k$ are generated in a quasirandom manner \cite{Frolov1995,Korobov1999}:
\begin{equation}
\alpha_k =
   \left[\left\lfloor\frac{1}{2}k(k+1)\sqrt{p_{\alpha}}\right\rfloor(A_2-A_1)+A_1\right]
   +i\left[\left\lfloor\frac{1}{2}k(k+1)\sqrt{q_{\alpha}}\right\rfloor(A'_2-A'_1)+A'_1\right]\,,
\end{equation}
where $\lfloor{x}\rfloor$ designates the fractional part of $x$, $p_{\alpha}$ and $q_{\alpha}$ are some prime numbers, and $[A_1,A_2]$ and $[A'_1,A'_2]$ are real variational intervals, which need to be optimized. Parameters $\beta_k$ and $\gamma_k$ are obtained in a similar way.

The bound state for the helium atom was calculated as in \cite{Korobov2000}, a set of intermediate states consist of a state with the total angular momentum $L'=1$ and the spatial parity $\pi=-1$. For the quasi-bound metastable states of the antiprotonic helium we use the Complex Coordinate Rotation method \cite{Ho1983}, numerical details of the calculations for the antiprotonic helium are given in \cite{Korobov2003}. For the $\bar{p}$He$^+$ atom in its initial state $(n,l)$, the intermediate states span over $L'=\{L,L\pm1\}$ with $\pi=-(-1)^L$.

In the CCR approach the coordinates of the dynamical system are rotated to some angle $\varphi$, parameter of the complex rotation: $r_{ij}\rightarrow r_{ij} e^{i\varphi}$. Under this transformation the Hamiltonian changes as a function of $\varphi$
\begin{equation}\label{rotHam}
H_{\varphi} = T e^{-2 i \varphi} + V e^{-i \varphi},
\end{equation}
where $T$ and $V$ are the kinetic energy and Coulomb potential operators. The continuum spectrum of $H_{\varphi}$ is rotated on the complex plane around branch points ("thresholds") to "uncover" resonant poles situated on the unphysical sheet of the Riemann surface. The resonance energy is then determined by solving the complex eigenvalue problem for the "rotated" Hamiltonian
\begin{equation}
(H_{\varphi} - E)\Psi_{\varphi} = 0, \label{roteqn}
\end{equation}
The eigenfunction $\Psi_{\varphi}$ obtained from Eq.~(\ref{roteqn}), is square-integrable and the corresponding complex eigenvalue $E = E_r - i\Gamma/2$ defines the energy $E_r$ and the width of the resonance, $\Gamma$, the latter is being related to the Auger rate as $\lambda_A = \Gamma/\hbar$.

\subsection{Static dipole polarizability}
\label{sec:dipole}

\begin{table}[b]
\caption{Test of convergence of the CCR calculations for the dipole scalar, $\alpha_s$, and tensor, $\alpha_t$, polarizability. The (31,30) state of $^4\mbox{He}^+\bar{p}\,$ atom. The last line is the Feshbach closed channel calculation. Here for simplicity we use the same number of basis functions $N$ for the initial and all ($L'=L,L\pm1$) intermediate states.}
\label{tab:polariz_31_30}
\begin{tabular}{c@{\hspace{6mm}}c@{\hspace{6mm}}c}
\hline\hline
$N$ & $\alpha_s$ & $\alpha_t\times10^{3}$ \\
\hline
2200 & $2.003\,118 + i\> 0.000\,531$ & $0.176\,288 - i\> 0.000\,441$ \\
3400 & $2.003\,011 + i\> 0.000\,424$ & $0.176\,341 - i\> 0.000\,342$ \\
4400 & $2.003\,030 + i\> 0.000\,455$ & $0.176\,332 - i\> 0.000\,384$ \\
5400 & $2.003\,033 + i\> 0.000\,453$ & $0.176\,330 - i\> 0.000\,382$ \\[1pt]
$\infty$&$2.003\,033(1) + i\> 0.000\,453$ & $0.176\,330(1) - i\> 0.000\,382$ \\[2pt]
\hline
1000 & 2.0031(4) & 0.1762(4) \\
\hline\hline
\end{tabular}
\end{table}

The static dipole polarizability tensor operator, which is a tensor of rank 2, on a subspace of fixed total angular momentum $L$ can be represented~\cite{Landau} by a scalar, $\alpha_s$, and irreducible tensor, $\alpha_t$, operators:
\begin{equation}\label{eq:tenor_polarizability}
\hat{\alpha}^{ij}_d(n,l) =
   \alpha_s(n,l)\,\delta_{ij}
   +\alpha_t(n,l)
   \left[
      \hat{L}_i\hat{L}_j+\hat{L}_j\hat{L}_i-\frac{2}{3}\delta_{ij}\hat{\mathbf{L}}^2
   \right]
\end{equation}
We use notation $\hat{o}\,$ to distinguish between operators and c-numbers.
The coefficients $\alpha_s$ and $\alpha_t$ then may be expressed in terms of three contributions corresponding to the possible values of the angular momentum of intermediate state, $L'=L,L\pm1$ (see Ref.~\cite{Schiller2014} for details)
\begin{subequations}\label{eq:alpha}
\begin{align}
  \alpha_s&=\frac{1}{3}\bigl(a_{L-1}+a_L+a_{L+1}\bigr)\,,\label{eq:a_s}\\[2mm]
  \alpha_t&=-\frac{a_{L-1}}{2L(2L\!-\!1)}+\frac{a_{L}}{2L(L\!+\!1)}-\frac{a_{L+1}}{2(L\!+\!1)(2L\!+\!3)}\,,
  \label{eq:a_t}
\end{align}
\end{subequations}
and $a_{L'}$ can be calculated by summing up the oscillator strengths as follows
\begin{equation}\label{eq:alphaL}
  a_{L'}=3\sum_{n_s}\frac{\bar{f}_{n_sn_0}^{(1)}(L',L)}{(E_{n_s}-E_{n_0})^2}\,,\qquad L'=L,L\pm1,
\end{equation}
where the 2$^l$-pole averaged oscillator strength is defined in terms of reduced matrix elements by the expression
\begin{equation}\label{eq:osc}
\bar{f}_{n_1n_2}^{(l)} = \frac{8\pi(E_{n_1}-E_{n_2})}{(2l+1)^2(2L+1)}\,
   \left|
   \left\langle
      \Psi_{n_1}\left\|\sum_i Z_ir_i^l\>Y_l^m\bigl(\hat{r}_i\bigr)\right\|\Psi_{n_2}
   \right\rangle
   \right|^2.
\end{equation}

For the CCR calculations we use a modified version of the perturbation theory  provided by the theorem \cite{Simon}.

\textbf{Theorem.\,} Let $H$ be a three-body Hamiltonian with Coulomb pairwise interaction, and $W(\theta)$ be a dilatation analytic "small" perturbation of a complex parameter $\theta$ (for the CCR we choose $\theta=i\varphi$). Let $E_0$ be an isolated simple resonance energy (discrete eigenvalue of $H(\theta)$). Then for $\beta$ small, there is exactly one eigenstate of $H(\theta)+\beta W(\theta)$ near $E_0$ and
\begin{equation}\label{perturb:expansion}
E(\beta)=E_0+a_1\beta+a_2\beta^2+\dots
\end{equation}
is analytic near $\beta=0$. In particular,
\begin{equation}
\begin{array}{@{}l}\displaystyle
a_1 = E'(0) =
   \left\langle\Psi^*_0(\theta)\left| W(\theta) \right|\Psi_0(\theta)\right\rangle,
\\[3mm]\displaystyle
a_2 = \sum_{n\ne0}
   \frac{
      \left\langle\Psi^*_0(\theta)\left| W(\theta) \right|\Psi_n(\theta)\right\rangle
      \left\langle\Psi^*_n(\theta)\left| W(\theta) \right|\Psi_0(\theta)\right\rangle}
         {E_0-E_n(\theta)}
\end{array}
\end{equation}
where the sum is carried out over the states of discrete and continuum spectra of the rotated Hamiltonian $H(\theta)$.

It is assumed that the wave functions are normalized as $\left\langle\Psi^*(\theta),\Psi(\theta)\right\rangle=1$. Coefficients $a_1$, $a_2$, etc do not depend on $\theta$ if only rotated branches of the continuum spectrum of $H(\theta)$ uncover $E_0$ and its vicinity on the complex plane. These coefficients are complex and the imaginary part contributes to the width of the resonance, to the imaginary part of the complex energy of Eq.~(\ref{roteqn}) $E = E_r - i\Gamma/2$, as it follows from Eq.~(\ref{perturb:expansion}).

In Table \ref{tab:polariz_31_30} convergence of the (31,30) state of $\bar{p}^4$He$^+$ is studied. This state is of much importance as a daughter state for the two-photon precision spectroscopy of the $(33,32)\to(31,30)$ transition.

\subsection{Dispersion coefficients}

The long-range interaction between two neutral atoms can be expanded in terms of a series of inverse powers of the separation distance $R$ \cite{Dalgarno1966,Dalgarno1967}:
\begin{equation}
U(R) = -\frac{C_6}{R^6}-\frac{C_8}{R^8}+\dots
\end{equation}
where $C_6$, $C_8$, etc, are the dispersion coefficients.
For two like atoms that are not both in their ground states, the perturbation theory of the dispersion coefficients has been well discussed in Refs.~\cite{Marinescu1995,Yan1996}. In the case of two different neutral atoms, the dispersion coefficients may be derived similarly in the frame of the perturbation theory.

Let us consider a dimer system composed of He in its ground state $(L_{\rm He}=0, M_{\rm He}=0)$ and $\bar{p}$He$^+$ in its excited state $(n, L, M)$. The electric interaction potential between He and $\bar{p}$He$^+$ at large separation $R$ can be expressed as a multipole harmonic expansion \cite{Marinescu1995,Yan1996,Zhang2004}
\begin{equation}\label{eq:VlL}
V(R;1,2) = \sum_{l_1=0}^{\infty}\sum_{l_2=0}^{\infty}\frac{V_{l_1l_2}}{R^{l_1+l_2+1}}\,,
\end{equation}
where
\begin{equation}
V(R;1,2) = \sum_{i,j=1}^3 \frac{Z_{1i}Z_{2j}}{\bigl|(\mathbf{r}_{2j}+\mathbf{R})-\mathbf{r}_{1i}\bigr|},
\end{equation}
and $\mathbf{r}_{1i}$ and $\mathbf{r}_{2j}$ are the center of mass position vectors of the three particles for the helium and antiprotonic helium atoms, respectively, $Z_{1i}$ and $Z_{2j}$ are the charges of particles for corresponding atoms.

If $z$ axis of frames of both atoms is taken along $\mathbf{R}$, expression for $V_{l_1l_2}$ may be written explicitly,
\begin{equation}\label{V_ll}
\begin{array}{@{}l}\displaystyle
V_{l_1l_2} = \sum_{m}
   \frac{(-1)^{l_2}(4\pi)(l_1+l_2)!}{\sqrt{(2l_1\!+\!1)(2l_2\!+\!1)(l_1\!-\!m)!(l_1\!+\!m)!(l_2\!-\!m)!(l_2\!+\!m)!}}
   \mathcal{M}_{l_1}^{[1]m}\mathcal{M}_{l_2}^{[2]-m}
\\[4mm]\displaystyle\hspace{15mm}
 = \frac{(-1)^{l_2}4\pi}{\sqrt{(2l_1\!+\!1)(2l_2\!+\!1)}}
   \left(\begin{matrix}2(l_1\!+\!l_2)\\2l_1\end{matrix}\right)^{\frac{1}{2}}
   \left\{\mathcal{M}_{l_1}^{[1]}\!\otimes\!\mathcal{M}_{l_2}^{[2]}\right\}_{l_1\!+\!l_2,0}\>,
\end{array}
\end{equation}
here $\mathcal{M}_L^m$ are the multipole moments of an atom:
\begin{equation}
\mathcal{M}_l^m = \sum_i\> Z_i r_i^l Y_l^m(\hat{\mathbf{r}}_i).
\end{equation}

In Eq.~(\ref{V_ll}) and in what follows superscripts in square brackets denote a particular subsystem (or atom), namely, 1 stands for the helium atom and 2 is for the antiprotonic helium.

Let $\Psi(1,2)$ be an eigenfunction of the interacting system:
\begin{equation}\label{main_eq}
\left[H_0+V(1,2)\right]\Psi(1,2) = E\,\Psi(1,2).
\end{equation}
Then assuming that $V(1,2)$ is small, one may use the following expansion
\begin{equation}\label{pert_expan}
\Psi(1,2) = \sum_{n=0}^\infty \Psi_n(1,2),
\qquad
E(1,2) = \sum_{n=0}^\infty E_n(1,2),
\end{equation}
where the zeroth-order wave function can be written as a product of two individual atomic wave functions
\begin{equation}
\Psi_0(1,2) = \psi_0^{[1]}\psi_0^{[2]},
\end{equation}
and the associated state energy is $E_0=E_0^{[1]}+E_0^{[2]}$.

\begin{table}[t]
\caption{\label{tab:phe4_alpha_c6_35}Test of convergence for the TRK sum rule, $\mathfrak{L}$, the static scalar dipole polarizability $\alpha_s$ and dispersion coefficient $C_6(M=0)$ for the (36,35) state of $\bar{p}^4$He$^+$ atom for a finite and infinite nuclear mass of He atom. $N_{L'}$ is the number of basis functions for the intermediate states with the total angular momentum $L'$.}
\begin{tabular}{ccc@{\hspace{7mm}}c@{\hspace{7mm}}c@{\hspace{6mm}}c@{\hspace{3mm}}c}
  \hline\hline
  &&&&& \multicolumn{2}{c}{$C_6(M=0)$} \\ \cline{6-7}
\vrule width 0pt height 10.5pt
  $N_{L-1}$  &  $~N_{L}~$ &  $N_{L+1}$ &
  \multicolumn{1}{c}{$\mathfrak{L}$~~~} & \multicolumn{1}{c}{$\alpha_s$~~~~~~} & \multicolumn{1}{c}{$^{\infty}$He$-\bar{p}^4$He$^+$~~~~} & \multicolumn{1}{c}{$^{4}$He$-\bar{p}^4$He$^+$} \\
 \hline
   140&   50&  130 & 1.008\,927\,6047 & 0.899\,415\,52  & 1.311\,880\,85 & 1.312\,410\,57\\
   300&  100&  300 & 1.001\,467\,4291 & 0.923\,357\,23  & 1.317\,432\,70 & 1.317\,964\,83\\
   500&  220&  500 & 1.001\,096\,8815 & 0.924\,043\,64  & 1.317\,652\,12 & 1.318\,184\,36\\
   700&  400&  700 & 1.001\,093\,2793 & 0.924\,050\,35  & 1.317\,652\,18 & 1.318\,184\,41\\
   900&  600&  900 & 1.001\,093\,0402 & 0.924\,050\,34  & 1.317\,654\,27 & 1.318\,186\,51\\
  1100&  800& 1100 & 1.001\,092\,9902 & 0.924\,050\,76  & 1.317\,654\,23 & 1.318\,186\,47\\
\multicolumn{3}{c}{$\mathfrak{R}$} & 1.001\,092\,9904 & & & \\
\multicolumn{3}{l}{Convergent values}&&0.924\,051(2) & 1.317\,654(1) & 1.318\,186(2)\\
   \hline\hline
\end{tabular}
\end{table}

\begin{table}[t]
\caption{\label{tab:phe4_alpha_c6_31}Test of convergence for the static scalar dipole polarizability $\alpha_s$ and dispersion coefficients $C_6(M=0)$ for the the (32,31) state of $\bar{p}^4$He$^+$ atom, for a finite and infinite nuclear mass of He atom. $N_{L'}$ is the number of basis functions for the intermediate states with the total angular momentum $L'$.}
\begin{tabular}{ccc@{\hspace{7mm}}c@{\hspace{6mm}}c@{\hspace{3mm}}c}
\hline\hline
  &&&& \multicolumn{2}{c}{$C_6(M=0)$} \\
\cline{5-6}
\vrule width 0pt height 10.5pt
  $N_{L-1}$  &  $N_{L}$ &  $N_{L+1}$ & \multicolumn{1}{c}{$\hspace*{-5mm}\alpha_s$}
& \multicolumn{1}{c}{$^{\infty}$He$-\bar{p}^4$He$^+$~~}
& \multicolumn{1}{c}{$^{4}$He$-\bar{p}^4$He$^+$~~} \\
 \hline
  1000& 1000 & 1000   & $0.362\,24+i\,0.013\,25$ & $1.617\,238+i\,0.000\,012$ & $1.617\,892+i\,0.000\,012$ \\
  1500& 1500 & 1500   & $0.352\,81+i\,0.011\,30$ & $1.617\,232+i\,0.000\,008$ & $1.617\,886+i\,0.000\,008$ \\
  2400& 2400 & 2400   & $0.352\,56+i\,0.011\,76$ & $1.617\,233+i\,0.000\,007$ & $1.617\,887+i\,0.000\,007$ \\
  3300& 3300 & 3300   & $0.352\,61+i\,0.011\,72$ & $1.617\,233+i\,0.000\,007$ & $1.617\,887+i\,0.000\,007$ \\
\hline
\multicolumn{3}{l}{Convergent values} & $0.352\,6(1)+i\,0.011\,8$ & $1.617\,233(1)+i\,0.000\,007$ & $1.617\,887(1)+i\,0.000\,007$ \\
 \hline\hline
\end{tabular}
\end{table}

Substituting expansions (\ref{pert_expan}) into Eq.~(\ref{main_eq}), one obtains a set of equations
\begin{equation}\label{pert}
\begin{array}{@{}l}\displaystyle
(H_0-E_0)\Psi_1+(V(1,2)-E_1)\Psi_0 = 0,
\\[2mm]\displaystyle
(H_0-E_0)\Psi_2+(V(1,2)-E_1)\Psi_1-E_2\Psi_0 = 0,
\\[1mm]\hspace{25mm}\cdots\cdots
\end{array}
\end{equation}
Equation (\ref{pert}) can be simplified by writing
\begin{equation}
\Psi_1 = \sum_{l_1=0}^\infty\sum_{l_2=0}^\infty \frac{\Omega_{l_1l_2}}{R^{l_1+l_2+1}}.
\end{equation}
where $\Omega_{l_1l_2}$ satisfies the equation
\begin{equation}
(H_0-E_0)\Omega_{l_1l_2}+(V_{l_1l_2}-\varepsilon_{l_1l_2}^{(1)})\Psi_0 = 0,
\end{equation}
with
\[
\varepsilon^{(1)}_{l_1l_2} = \left\langle\Psi_0|V_{l_1l_2}|\Psi_0\right\rangle.
\]

For neutral atoms, of which one is in the ground $S$ state, the first-order energy vanishes, $E_1=0$ \cite{Yan1996}. The second order term is then expressed
\begin{equation}
E_2(1,2) = -\sum_{L_1=1}^\infty\sum_{L_2=1}^\infty \frac{\varepsilon_{l_1l_2}^{(2)}}{R^{2(l_1+l_2+1)}},
\end{equation}
and
\begin{equation}\label{eps2}
\varepsilon^{(2)}_{l_1l_2} = \left\langle \Psi_0 | V_{l_1l_2} | \Omega_{l_1l_2} \right\rangle
\end{equation}
Now the dispersion coefficient $C_6$ may be written explicitly as
\begin{equation}\label{eq:C6_Mb}
\begin{array}{@{}l}\displaystyle
C_6 = \varepsilon^{(2)}_{11}
 = \sum_{ij} \frac{\left|\left\langle00\big|V_{11}\big|ij\right\rangle\right|^2}{E_{ij}-E_{00}}
\\[5mm]\displaystyle\hspace{7mm}
 = \frac{1}{3}\left(\frac{4\pi}{3}\right)^2
   \left\{
   \sum_{i,n}
   \frac{
         \left\langle 0 \left\|\mathcal{M}_1^{[1]}\right\| i \right\rangle^2
         \left\langle 0L \left\|\mathcal{M}_1^{[2]}\right\| n(L\!+\!1) \right\rangle^2}{E_{ij}-E_{00}}\;
      \frac{(L+1)(5L+6)-3M^2}{(L+1)(2L+1)(2L+3)}
   \right.
\\[3mm]\displaystyle\hspace{27mm}
   +\sum_{i,n}
   \frac{
         \left\langle 0 \left\|\mathcal{M}_1^{[1]}\right\| i \right\rangle^2
         \left\langle 0L \left\|\mathcal{M}_1^{[2]}\right\| nL \right\rangle^2}{E_{ij}-E_{00}}\;
      \frac{L(L+1)+3M^2}{L(L+1)(2L+1)}
\\[3mm]\displaystyle\hspace{27mm}
   +\left.\sum_{i,n}
   \frac{
         \left\langle 0 \left\|\mathcal{M}_1^{[1]}\right\| i \right\rangle^2
         \left\langle 0L \left\|\mathcal{M}_1^{[2]}\right\| n(L\!-\!1) \right\rangle^2}{E_{ij}-E_{00}}\;
      \frac{L(5L-1)-3M^2}{L(2L-1)(2L+1)}
   \right\}
\end{array}
\end{equation}
here $L$ is the total orbital angular momentum of the antiprotonic helium state, while index $i$ runs over the $P$ states of the helium atom.

\begin{table}[b]
\caption{\label{tab:alpha} The static dipole polarizability, $\alpha_s$ and $\alpha_t$, for metastable states ($n$,$L$) of $\bar{p}^3$He$^+$ and $\bar{p}^4$He$^+$.}
\begin{tabular}{c@{\hspace{7mm}}r@{\hspace{3mm}}r@{\hspace{7mm}}r@{\hspace{3mm}}r}
\hline\hline
\vrule width 0pt height 10.5pt
  & \multicolumn{2}{c}{$\bar{p}^3$He$^+$} & \multicolumn{2}{c}{$\bar{p}^4$He$^+$}\\
\cline{2-3} \cline{4-5}
\vrule width 0pt height 10.5pt
 ($n$,$L$) & \multicolumn{1}{c}{$\alpha_s$} & \multicolumn{1}{c}{$\alpha_t\times10^3$}
           & \multicolumn{1}{c}{$\alpha_s$} & \multicolumn{1}{c}{$~~\alpha_t\times10^3$} \\
\hline
    (31,30) &  $1.78619-i\,0.00081$ &  $0.21466+i\,0.00028$ &  $2.00303+i\,0.00045$ &  $0.17633-i\,0.00038$ \\[3pt]
    (32,31) &  1.56805~~~~~~~~      &  0.23644~~~~~~~~      &  $0.35261+i\,0.01172$ &  $1.13825-i\,0.00951$ \\[3pt]
    (33,31) &  $1.11874+i\,0.17764$ &  $2.16493-i\,0.13634$ &       ---~~~~~~~~~~~  &        ---~~~~~~~~~~~ \\
    (33,32) &  1.34708~~~~~~~~      &  0.26348~~~~~~~~      &  1.57429~~~~~~~~      &  0.22059~~~~~~~~      \\[3pt]
    (34,32) &  $1.10917+i\,0.00007$ &  $0.34856-i\,0.00005$ &  $1.34456+i\,0.00059$ &  $0.28779-i\,0.00045$ \\
    (34,33) &  1.12356~~~~~~~~      &  0.29495~~~~~~~~      &  1.36108~~~~~~~~      &  0.24470~~~~~~~~      \\[3pt]
    (35,32) &  $0.89521+i\,0.00735$ &  $0.44815-i\,0.00546$ &  $2.14937+i\,0.06382$ & $-0.22096-i\,0.04261$ \\
    (35,33) &  0.87871~~~~~~~~      &  0.39529~~~~~~~~      &  1.12525~~~~~~~~      &  0.32360~~~~~~~~      \\
    (35,34) &  0.89434~~~~~~~~      &  0.33300~~~~~~~~      &  1.14520~~~~~~~~      & 0.27278~~~~~~~~      \\[3pt]
    (36,32) &  $1.37921+i\,0.07366$ &  $1.03868-i\,0.04512$ &  $1.94467+i\,0.17316$ & $-0.07879-i\,0.07193$\\
    (36,33) &  $0.70924+i\,0.00289$ &  $0.48140-i\,0.00197$ &  $0.90508+i\,0.00179$ & $0.42010-i\,0.00126$ \\
    (36,34) &  0.64656~~~~~~~~      &  0.44771~~~~~~~~      &  0.90739~~~~~~~~      & 0.36234~~~~~~~~      \\[3pt]
%    (36,35) & 0.6551774(2)&0.380156(1)  & 0.924051(2)  &0.306519(2) \\
    (37,33) &  $1.06192+i\,0.00487$ &  $0.24605-i\,0.00344$ &  $52.0183+i\,12.5824$ & $-31.8571-i\,8.3973$ \\
    (37,34) &  0.40244~~~~~~~~      &  0.59725~~~~~~~~      &  0.67456~~~~~~~~      & 0.47854~~~~~~~~      \\
    (37,35) &  0.40318~~~~~~~~      &  0.51175~~~~~~~~      &  0.68389~~~~~~~~      & 0.40841~~~~~~~~      \\[3pt]
%    (37,36) & 0.3997737(1)&0.4398256(1) & 0.6940326(2) & 0.3479884(2)  \\
    (38,33) &  $1.33938+i\,0.01240$ &  $0.14569-i\,0.00780$ &  $0.31099-i\,0.11508$ & $0.91666-i\,0.01162$ \\
    (38,34) &  $0.16689+i\,0.00029$ &  $0.78694-i\,0.00019$ &  $0.45394+i\,0.00207$ & $0.62294-i\,0.00138$ \\
    (38,35) &  0.15570~~~~~~~~      &  0.68149~~~~~~~~      &  0.44716~~~~~~~~      &    0.54035~~~~~~~~   \\[3pt]
%    (38,37) & 0.118665(1) &0.5167323(1) &  0.4497719(1) &0.39994185(2)  \\
    (39,34) &  $0.02278-i\,0.07504$ &  $0.98716+i\,0.04701$ &  $0.24206+i\,0.00157$ & $0.80977-i\,0.00095$ \\
    (39,35) &        ---~~~~~~~~~~~ &       ---~~~~~~~~~~~  &   0.21561~~~~~~~~     &    0.70833~~~~~~~~   \\[3pt]
    (40,35) & $-$0.30463~~~~~~~~    &  1.16374~~~~~~~~      & $-0.51572+i\,0.11763$ & $1.29819-i\,0.05401$ \\
    (40,36) &        ---~~~~~~~~~~~ &       ---~~~~~~~~~~~  & $-$0.02227~~~~~~~~    & 0.80180~~~~~~~~      \\[3pt]
    (41,35) &        ---~~~~~~~~~~~ &       ---~~~~~~~~~~~  & $-1.92803+i\,0.00669$ & $2.20173-i\,0.00416$ \\
    \hline\hline
  \end{tabular}
\end{table}

\section{Calculation and results}

\begin{table}[t]
\caption{\label{tab:c6-m}Dispersion coefficients $C_6(M)$ for long-range interaction between He and $\bar{p}$He$^+$.}
\begin{tabular}{c@{\hspace{6mm}}l@{\hspace{3mm}}lc@{\hspace{6mm}}l@{\hspace{3mm}}l}
\hline\hline
\vrule width 0pt height 10.5pt depth 3pt
 & \multicolumn{2}{c}{$^3$He$-\bar{p}^3$He$^+$} && \multicolumn{2}{c}{$^4$He$-\bar{p}^4$He$^+$} \\
 \cline{2-3}\cline{5-6}
\vrule width 0pt height 10.5pt depth 3.5pt
 $(n,L)$ & $M=0$ & $M=\pm{L}$ &  & $M=0$ & $M=\pm{L}$ \\
\hline
   (31,30) & 1.612\,77   & 1.619\,74   && 1.701\,107  & 1.712\,912  \\
   (32,31) & 1.525\,611  & 1.524\,632  && 1.617\,89   & 1.622\,28   \\
   (33,31) & 1.613\,7    & 1.469\,8    && 1.531\,20   & 1.517\,26   \\
   (33,32) & 1.445\,058  & 1.432\,646  && 1.527\,723  & 1.527\,045  \\
   (34,32) & 1.387\,246  & 1.341\,281  && 1.455\,162  & 1.429\,245  \\
   (34,33) & 1.372\,941  & 1.344\,852  && 1.449\,535  & 1.437\,956  \\
   (35,32) & 1.359\,370  & 1.268\,273  && 1.453\,     & 1.375\,     \\
   (35,33) & 1.335\,233  & 1.264\,855  && 1.391\,356  & 1.348\,092  \\
   (35,34) & 1.311\,625  & 1.262\,614  && 1.378\,966  & 1.352\,603  \\
   (36,32) & 1.467\,     & 1.281\,     && 1.456\,     & 1.328\,     \\
   (36,33) & 1.331\,77   & 1.205\,42   && 1.360\,841  & 1.275\,199  \\
   (36,34) & 1.298\,994  & 1.196\,786  && 1.339\,492  & 1.273\,340  \\
   (36,35) & 1.264\,137  & 1.187\,617  && 1.318\,187  & 1.272\,250  \\
   (37,33) & 1.361\,113  & 1.163\,439  && 2.135\,     & 1.698\,     \\
   (37,34) & 1.324\,189  & 1.153\,346  && 1.331\,131  & 1.212\,822  \\
   (37,35) & 1.281\,479  & 1.138\,709  && 1.301\,665  & 1.206\,030  \\
   (37,36) & 1.234\,243  & 1.121\,883  && 1.269\,911  & 1.198\,418  \\
   (38,33) & 1.443\,6    & 1.154\,0    && 1.431\,     & 1.212\,     \\
   (38,34) & 1.390\,551  & 1.133\,412  && 1.357\,03   & 1.172\,43   \\
   (38,35) & 1.341\,39   & 1.114\,32   && 1.320\,232  & 1.160\,541  \\
   (39,34) & 1.500\,6    & 1.137\,8    && 1.421\,71   & 1.154\,37   \\
   (39,35) & 1.446\,4    & 1.115\,1    && 1.376\,708  & 1.136\,844  \\
   (40,35) & 1.598\,173  & 1.141\,394  && 1.480\,3    & 1.139\,5    \\
\hline \hline
\end{tabular}
\end{table}

Using Eqs.~(\ref{eq:alpha})--(\ref{eq:alphaL}), one can get the static dipole polarizability (scalar, $\alpha_s$, and tensor part, $\alpha_t$) for metastable states of $\bar{p}$He$^+$. In order to check the validity of our calculations, we use the generalized Thomas-Reiche-Kuhn (TRK) sum rule for the oscillator strengths developed by Z.-C. Yan and co-workers \cite{Zhou2006}. For $\bar{p}^4$He$^+$, the left hand side $\mathfrak{L}$ and right hand side $\mathfrak{R}$ of the TRK equality are expressed as follows,
\begin{equation}
  \mathfrak{L}=\sum_{n_t}\bar{f}_{n_tn_b}^{(1)}\,,
\qquad
  \mathfrak{R}=\frac{4}{m_{^4{\rm{He}}^{2+}}}+\frac{1}{m_{\bar{p}}}+1\,.
\end{equation}
A test of convergence of $\mathfrak{L}$ and a comparison with the exact value, $\mathfrak{R}$, for the (36,35) state of $\bar{p}^4$He$^+$ are listed in Table \ref{tab:phe4_alpha_c6_35}, what demonstrates reliability of our calculations. In the numerical results of this section the CODATA10 \cite{CODATA10} recommended values were adopted.

Table \ref{tab:phe4_alpha_c6_35} and Table \ref{tab:phe4_alpha_c6_31} provide tests of convergence of $\alpha_s$ and $C_6(M=0)$ for the two kind of states: the metastable (36,35) state $\bar{p}^4$He$^+$ decaying via slow radiative transition, and the (32,31) state $\bar{p}^4$He$^+$, where Auger rate becomes dominant. Appearance of the imaginary part in the data and the physical meaning of it has been clarified in Sec.~\ref{sec:dipole} right after the Theorem. As is seen from Table \ref{tab:phe4_alpha_c6_31}, the imaginary part is substantially large in the case of polarizability of the (31,30) state. That may be explained by strong correlation with broad short-lived states having excited electronic configuration and lying in a vicinity of the (31,30) state on the Reimann surface of complex energy (see \cite{Yamazaki2002}, Sec.~4.8 and discussion below). It is also important to note that the dispersion coefficients, $C_6(M)$, are much less affected by this phenomenon and the imaginary part may be ignored. Since the finite mass of a helium nucleus produce a visible effect on $C_6(M)$ we explicitly compare the $C_6(M)$ for $^4$He--$\bar{p}^4$He$^+$ and $^{\infty}$He--$\bar{p}^4$He$^+$ in the Tables. A choice of a finite or infinite mass for the helium atom causes changes in the fourth decimal place in the $C_6$ coefficient. That is essential effect and in our final calculations of the dispersion coefficients we use the wave functions for the helium atom obtained with the finite mass of a nucleus.

Numerical calculations of the dipole polarizabilities ($\alpha_s$ and $\alpha_t$) for metastable states in the antiprotonic helium  are presented in Table \ref{tab:alpha}. Results were obtained with the use of the CCR method, thus the final values have an imaginary part. Some of the states, particularly the (37,33) state in $\bar{p}^4$He$^+$, have anomalously large polarizability, which makes the states unstable against collisions. Such anomalous behaviour is again connected with the excited electron "Rydberg" states, which strongly affects the overall polarizability of the "atom". Yet another, maybe less obvious example, the (32,31) state in $\bar{p}^4$He$^+$ atom. This state, as it was observed in experiment \cite{Yamaguchi2002}, at large densities become unstable. The most apparent explanation of this phenomena is considerable (by a factor of 10) increase of the tensor polarizability. The small value of $\alpha_t$ should not be deceitful, there is a large prefactor for this contribution: $2M^2-(2/3)L(L\!+\!1)\sim2\cdot10^3$, see Eq.~(\ref{eq:tenor_polarizability}).

In Table \ref{tab:c6-m} the main result of this work, the dispersion coefficients $C_6(M)$, are presented. We expect that all the digits indicated are significant. As can be concluded from Eq.~(\ref{eq:C6_Mb}), the value of $C_6(M)$ depends on the magnetic quantum number $M$ of the antiprotonic helium state as:
\begin{equation}
C_6(M) = C_6 + D_6\,M^2,
\end{equation}
where $C_6$ and $D_6$ are some coefficients that may be obtained from $C_6(M=0)$ and $C_6(M=\pm L)$ as follows:
\begin{equation}
C_6 = C_6(0), \qquad D_6 = \bigl[C_6(L)-C_6(0)\bigr]/L^2.
\end{equation}
Thus only numerical values for cases $M=0$ and $M=\pm L$ are displayed in the Table. One may find that dependence on $M$ of $C_6(M)$ increases with $n$ and $L$. That may be explained by growing asymmetry of the state: antiproton become to spend more time outside of the electronic cloud. The data in the Table demonstrates quite a regular behaviour of $C_6$ coefficients with one exception: the (37,35) state in $\bar{p}^4$He$^+$ atom has larger values of $C_6$. That means that atoms in this state experience stronger attraction while interacting with the helium atoms of the target, and eventually stronger collisional disintegration of the atom and subsequent prompt annihilation of $\bar{p}$.

Using the numerical values of $C_6(M)$ one can get a rough estimate for the collisional shift for transitions between metastable states of the antiprotonic helium. Applying the simplest "static" approximation, the frequency shift can be expressed as
\begin{equation}
  \Delta\omega=\frac{C_i-C_f}{R_{av}^6}\,,
\end{equation}
where $C_i$ and $C_f$ are the dispersion coefficients $C_6(M)$ for the initial and final states, respectively, and $R_{av}$ is an averaged separation distance between atoms. Assuming the separation distance $R\approx19$~a.u. that approximately corresponds to a number density of $10^{21}$ cm$^{-1}$, the spectroscopy shift for the (33,32)$\to$(32,31) transition would be 12 GHz. More sophisticated models of the collisional shift and broadening may be found in \cite{demtroeder,sobelman,allard}.

In conclusion, the static dipole polarizability and the leading order van Der Waals coefficient $C_6$ have been evaluated for metastable states of $\bar{p}^3$He$^+$ and $\bar{p}^4$He$^+$ atoms for a wide range of metastable states of practical interest. These data may be used for estimating the collisional shift and broadening of the transition frequencies of the antiprotonic helium at low temperature.

\begin{acknowledgments}
Authors thank Z.-C. Yan, T.-Y. Shi and L.-Y. Tang for their helpful discussion. This work was supported by the National Basic Research Program of China (973 Program) under Grant No. 2010CB832803, by the NSFC under Grants No. 11004221 and No. 11274348. V.I.K.\ also acknowledges support of the Russian Foundation for Basic Research under Grant No.~15-02-01906-a.
\end{acknowledgments}

\end{document}